\documentclass[aps,pre,twocolumn,groupedaddress]{revtex4}
\usepackage{graphicx}
\def\be{\begin{equation}}
\def\ee{\end{equation}}
\begin{document}
\draft

\title{Thermodynamics of driven collisionless systems}
\author{Felipe B. Rizzato\footnote{rizzato@if.ufrgs.br}, Renato Pakter\footnote{pakter@if.ufrgs.br}, 
and Yan Levin\footnote{levin@if.ufrgs.br}}
\address{Instituto de F\'{\i}sica,
Universidade Federal do Rio Grande do Sul\\
Caixa Postal 15051, 91501-970, Porto Alegre, RS, Brazil
}
\begin{abstract}
A statistical theory is presented which allows to calculate the stationary state achieved by a driven system
after a  process of collisionless relaxation.  The theory is applied to study an 
electron beam driven by an external electric field. The Vlasov equation with appropriate 
boundary conditions is solved analytically and compared with the molecular dynamics simulation. 
A perfect agreement is found between the theory and the simulations. 
The full current-voltage phase diagram
is constructed.   
\end{abstract}
%
\maketitle
%
\section{Introduction}
Unlike the equilibrium thermodynamics and statistical mechanics, which are well developed after the 
pioneering works of Boltzmann and Gibbs, our understanding of non-equilibrium thermodynamics is restricted to some special
models and cases.  Stochastic lattice gases have provided a fertile testing ground for studying
non-equilibrium stationary states in driven systems~\cite{MaDi99,derrida92,derrida93}.  These models exhibit a variety of phase transition arising
from a diffusive (collisional) relaxation. 
For some of these models local equilibrium and hydrodynamic equations have been
derived rigorously~\cite{BeSo07}.

There are, however, other physical systems for which the approach to final
stationary state is through a process of collisionless relaxation~\cite{Ly67,Pa90,ChSo98,Ch06,AnFa07,levin08,LePa08}.  
Gravitational systems and confined one component plasmas 
are just two such examples.  For these systems the collision duration time diverges and the relaxation 
is governed by the collisionless Boltzmann (Vlasov) equation~\cite{davidson01}. In the thermodynamic limit, 
the collisionless relaxation process leads to 
non-Maxwell-Boltzmann velocity distributions, even for stationary states without macroscopic currents. 
Unlike normal thermodynamic equilibrium, the stationary state which follows the collisionless relaxation depends explicitly on
the initial distribution of particle positions and velocities.  In spite of this complication, 
it was recently shown that it is possible to construct a statistical theory that quantitatively 
describes these states~\cite{levin08,LePa08}.

Beams of electrons driven by accelerating vacuum 
devices, like the thermionic valves, diodes, and magnetrons,  also  do not relax to 
the Maxwell-Boltzmann distribution~\cite{reiser94}. 
Unlike the driven stochastic lattice gases, these systems, however, are intrinsically collisionless.
An important practical question concerns the kinetic temperature distribution 
in thermionic devices in which the directed velocity  produced by 
the electric field is comparable to the thermal velocity~\cite{chip06}. 
This is particularly the case for the transitional 
region between Child-Langmuir and no-cutoff regimes in magnetrons, where the electric potential becomes 
comparable to the thermal energy~\cite{chip06}. Even when the final directed velocity is larger than the 
thermal velocity, there is a  region near the emitting cathode where thermal effects are important. It is of great 
practical 
interest to determine the extent of these regions \cite{diode1,diode2}.  Furthermore, since 
in these systems the collision duration time diverges, there is no local equilibrium,  and one can not 
{\it a priori} postulate an equation of state relating the 
beam density and the beam temperature, as for adiabatic or isothermal processes. Instead, 
given the properties of thermionic filaments --- such as say the velocity distribution of the emitted electrons --- one should 
solve the boundary value problem posed by the Vlasov equation.  The purpose of this
Letter is to develop a theoretical framework which will allow us to study the relaxation dynamics and the 
stationary states of collisionless driven systems.



As a prototype of a collisionless driven system, we
consider a beam of electrons, accelerated by an external electric field,  
traveling from an emitting (planar) cathode to a collecting (planar) anode across the device gap. 
The cathode, located at position $x=0$, is kept at 
electrostatic potential $\varphi(x=0) = 0$ and is heated to temperature 
$T_c$, resulting in the emission of electrons. 
After traversing the device gap, these electrons are 
collected at the  
cold anode ($T_a \approx 0$) located at $x=L$ and kept at 
potential $\varphi(x=L) = V>0$.  During the steady state operation, 
the region between the cathode and anode contains a total of $N$ electrons, resulting in 
a current density $j$.  Our goal is to relate $j$, to the potential difference $V$, the
number of electrons $N$, the device width $L$, and the cathode temperature $T_c$.    

For planar electrodes, 
particle distribution transverse to the $x$ axis can be taken to be uniform.  Furthermore,
the one particle distribution function for a collisionless system in a steady state must
satisfy the stationary Vlasov  equation,
\begin{equation}
v {\partial f \over \partial x} + {e \over m} \nabla \varphi(x) \, {\partial f \over \partial v}=0,
\label{equa1}
\end{equation}
where $e$ is the elementary charge, $m$ is the electron mass, and $f=f(x,v)$ is the static distribution function. In the
thermodynamic limit, Vlasov equation becomes exact for particles interacting by long range potentials~\cite{Br77}. 

It can 
be readily seen that the distribution functions of the form $f(x,v)=f[\varepsilon(x,v)]$, where $\varepsilon $ is 
the mean particle energy, 
$\varepsilon \equiv mv^2/2 - e \varphi(x)$, satisfy Eq. (\ref{equa1}). Therefore, 
if $f(\varepsilon)$  is specified at $x=0$, $f$ is then also determined for any other position, provided that
the electrostatic potential $\varphi(x)$ is known. This potential can, in turn, be calculated 
self-consistently from the solution of the 
Poisson equation 
\begin{equation}
\frac{d^2\varphi(x)}{d\,x^2} = {e \over \epsilon_0} \> n(x),
\label{equa2}
\end{equation}
where the particle density $n(x)$ is given by $n(x) = \int_v  f(\varepsilon) \, dv = \int_v f(mv^2/2 - e \varphi(x))\,dv$, 
the total particle number is $N=A \int_v \int_0^L f \, dx \, dv$, and the transverse cross sectional area 
of the essentially 1D device is $A$.  To represent both the thermal distribution near 
the cathode, and the fact that only particles with positive velocities actually move into the device gap, we choose at $x=0$ a 
unidirectional Maxwellian distribution of the form 
\begin{equation}
f (x=0,v) = \> \cases{ n_0 \sqrt{2 m \over \pi k_B T_c} \, \exp\left(-{ m\,v^2 \over 2 k_B T_c}\right) \>\> {\rm if} \>\> v \geq 0, \cr 0 \>\> {\rm if} \>\>v<0.}
\label{equa3}
\end{equation}
where $k_B$ is the Boltzmann constant, $T_c$ is the cathode temperature, and $n_0$ 
is the beam density at the cathode after the stationary state is achieved.  The value of $n_0$ can only be obtained
once the full problem has been resolved.  The distribution function over the length of the whole diode is then
\begin{widetext}
\begin{equation}
f (x,v) = \> \cases{ n_0 \sqrt{2 m \over \pi k_B T_c} \, \exp\left(-\frac{ m\,v^2}{ 2 k_B T_c}+\frac{e \varphi(x)}{k_B T_c}\right) \>{\rm if} \> v \geq v_{min}(x), \cr 0 \>\> {\rm if} \>\>v<v_{min}(x).}
\label{equa3}
\end{equation}
\end{widetext}
where $v_{min}(x)=\sqrt{2 e \varphi(x) \over m}$.

Integrating the distribution function $f[\varepsilon(x,v)]$ over the possible
values of velocity, we arrive at a nonlinear integro-differential equation for the electrostatic potential,
\begin{equation}
\frac{d^2 \varphi}{d\,x^2}={N e \over \varepsilon_0 A} {e^{{e \varphi(x)\over k_B T_c}\,}{\rm Erfc}\left( \sqrt{e \varphi(x)\over k_B T_c} \right) 
\over \int e^{{e \varphi(x)\over k_B T_c}\,}{\rm Erfc}\left( \sqrt{e \varphi(x)\over k_B T_c} \right) dx}.
\label{equa4}
\end{equation}
It is important to note the difference between this equation and the  Poisson-Boltzmann equation obtained  for usual collisional 
plasmas and electrolytes in the mean-field limit~\cite{Le02}.  Equation (\ref{equa4}) can be solved numerically, 
to yield the electrostatic potential and the distribution function for the electron beam in
the stationary state.

For systems with long range interactions, Vlasov equation should become exact 
in the thermodynamic limit.  
To confirm this for our system, we have performed molecular dynamics simulation of
an equivalent one dimensional model.  The simulated system 
consists of $N_s$ mutually interacting charged sheets of area $A$ --- each containing $n_s$ electrons of 
the same velocity --- 
moving along the $x$ axis, 
under the action of the external electric field produced by the grounded cathode $\varphi(0)=0$ and 
an anode kept at a fixed potential $\varphi(L)=V$. The interaction potential between the two sheets $G(x_i,x_j)$ is 
the Green's function~\cite{jack99} of the Laplace equation, $d^2G(x,y)/dx^2 =1/L\> \delta (x-y)$ with the 
boundary conditions $G(x=0,y)=G(x=L,y)=0$.
Solving this equation we obtain $G(x_i,x_j) = x_{_<}/L \> \big(x_{_>}/L-1\big)$, where $x_{<}$ and $x_{>}$  
are the smaller and the larger of the two particle coordinates $x_i$ and $x_j$.
The effective Hamiltonian for the sheet dynamics is then
\begin{equation}
H=\sum_i \left({m_s v_i^2 \over 2} - {e_s V \over L}  x_i  \right) - 
{1 \over 2}\>{e_s^2 L \over \varepsilon_0 A } \> \sum_{i,j} G(x_i,x_j),
\label{seraesta}
\end{equation}
%
where $e_s=n_s e$ and $m_s=n_s m$ are the charge and mass of each sheet respectively.  The acceleration of 
each simulated sheet then follows from the canonical equations of motion,
\begin{widetext}
\begin{equation}
\dot v= {e V\over m L} + {N e^2 \over 2 \varepsilon_0 m A } \left[ \left( {n^{[left]} - n^{[right]} \over N_s}\right) - 
\left(1-2\,{\overline{x} \over L}\right) \right],
\label{equa5}
\end{equation}
\end{widetext}
where $n^{[left(right)]}$ is the number of sheets to the left(right) of the one considered, and $\overline x$ denotes the 
positional average $\overline x = \sum_j x_j/N_s$. 
Since $0 \leq \overline x \leq L$, from Eq. (\ref{equa5}) one sees that the electron acceleration at 
the device entrance where $n^{[left]} \rightarrow 0$ and $n^{[right]} \rightarrow N_s$
satisfies $eV/mL - N e^2/2 \varepsilon_0 m A< \dot v (x=0) < eV/mL$, which reveals that in {\it space-charge dominated} 
devices where $eV/mL < N e^2/2 \varepsilon_0 m A$, acceleration at beam entrance may be zero or even negative \cite{pak01}. 
When the acceleration vanishes, the associated current is denoted as the {\it limiting } one. 
Since we wish to describe a hot cathode and a cycling current inside the device, we adopt the following 
strategy. We advance the simulation in small time steps, always obeying Eq. (\ref{equa5}). Whenever a particle crosses the anode 
and exits the system, it is re-injected at the cathode position. At this point all the particles in a small 
region $\delta_s$ around the cathode 
are re-thermalized, so as to ensure that the distribution there keeps its original form of a truncated Maxwellian. 
The width $\delta_s$ must be sufficiently small, $\delta_s \ll L$, but apart from this condition its precise value is 
arbitrary. The simulations were performed with $\delta_s/L=0.01$ and $N_s=50,000$. In all cases we start with 
a uniform distribution of sheets and compute the observables only after the system reaches its final stationary state.

To compare the predictions of the theory with the results of the simulations, we consider the density and the 
temperature distributions inside the diode.  The kinetic temperature is defined as 
\begin{equation}
k_B \, T(x)=\overline{v^2}(x) - \overline v^2(x).
\label{equa6}
\end{equation}
where the over-bar denotes the velocity average at a given position $x$. 
The theoretical averages are calculated using the 
distribution function $f[\varepsilon(x,v)]$, while in the
simulations, the averages are performed over the particle velocities within narrow bins along the $x$ axis. Note that
because of the asymmetry of the velocity distribution at $x=0$, $T(0)\ne T_c$.

It is convenient to scale space and time with the diode length 
$L$ and the plasma frequency $\omega_p^2 \equiv N e^2/ \epsilon_0 m L A$, 
respectively. Dimensionless coordinate and velocity can then be defined as 
$x^* = x/L$ and $v^*=v  / L \omega_p$. In addition, Eqs. (\ref{equa5}) and (\ref{equa3})
show that adimensional voltage and adimensional temperature can be defined 
as $V^* = e V/m L^2 \omega_p^2$ and $T^*= k_B T/m L^2 \omega_p^2$, 
respectively, and serve as the control parameters for the system. 

\begin{figure} [h!]
\vspace{0.5cm}
\begin{centering}
\includegraphics[width=8cm,height=9cm]{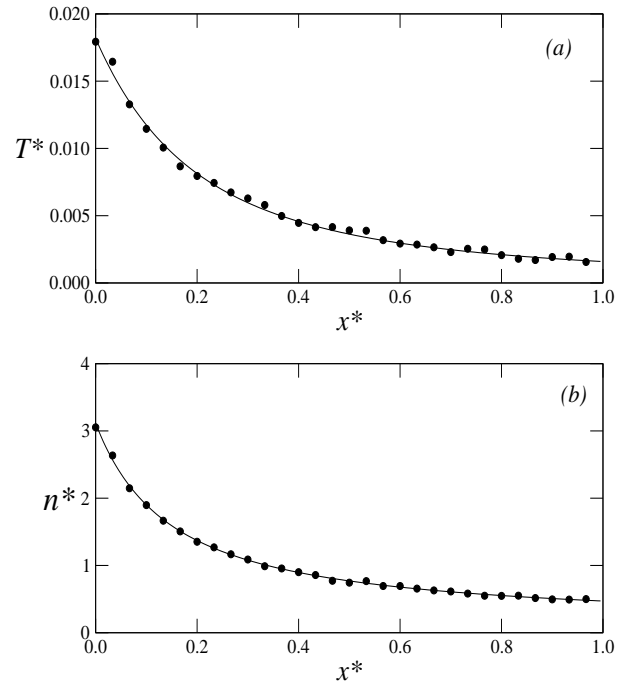}
\caption{(a) Temperature and (b) density versus position in the case of the limiting current and  $T_c^*=0.05$. Solid lines
represent the theoretical results, while the circles are the results of the simulations.\label{fig1.eps}}
\end{centering}
\end{figure}

In Fig. \ref{fig1.eps}(a) the scaled temperature $T^*$  
is plotted against the scaled coordinate $x^*$. We consider $T_c^* = 0.05$ and also 
consider a device operating at its limiting current, $\dot v (x=0) = 0$. A striking feature of this plot is that 
the temperature drops rapidly as one 
moves away from cathode towards anode. We next study the dependence of scaled density 
$n^*(x^*) = n(x) A\, L/N$ along the length of the diode. The density is very high near the cathode, 
where the average velocity is small.   It then drops rapidly towards the anode, where particles 
are accelerated up to high speeds,
see Fig. \ref{fig1.eps}(b). 
Agreement between the  simulations and the theory for both the kinetic temperature and density is excellent. 

\begin{figure} [h!]
\vspace{0.5cm}
\begin{centering}
\includegraphics[width=8cm,height=7cm]{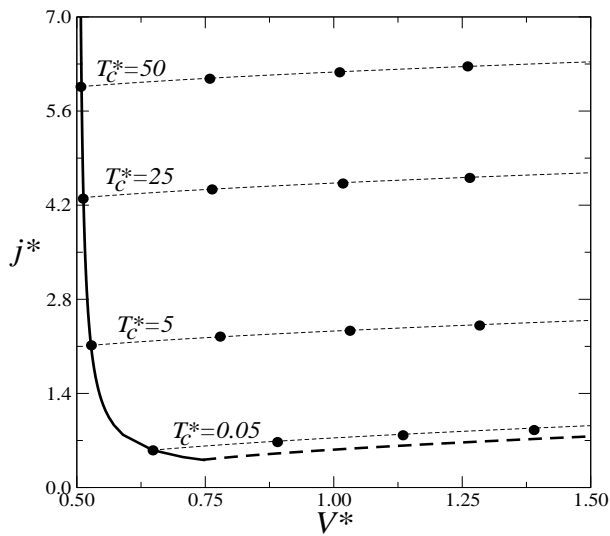}
\caption{Characteristic curves of  $j^*$  vs. $ V^*$. The thick solid line  represents the limiting current and the 
thick dashed line, the zero temperature limit. To the left of the solid curve, charge must be expelled from the 
system before a stationary state can be achieved.  Dotted lines represent the theoretical results for the indicated 
temperatures. The circles are the results of the simulations at the same temperatures.\label{fig2.eps}}
\end{centering}
\end{figure}

We now study the current-voltage phase diagram of the device.  
In general, current is a function of the voltage drop, the temperature, the gap length, and the total charge of
the device.  However, by measuring the time in units of one over the plasma frequency $\omega_p^{-1}$, and the length in units of the 
gap length $L$, we can scale away two of these variables. 
The current density can be calculated using 
\begin{equation}
j= - \, e \int v \, f \, dv.
\label{primeiraj}
\end{equation}
Since in the steady state the current does not depend on either time or
coordinate, integration along the $x$ axis and over the cross sectional area $A$ yields
\begin{equation}
j LA= - \, e \int v \, f \, dv dx d^2r_\perp.
\label{equa7}
\end{equation}
Furthermore, since the current density is measured in units  $[j] \sim [Ne v/A L]$, rescaling it in terms of the gap length and the plasma frequency,
we can write the reduced current density as  $j^*=-e j / \varepsilon_0 mL \omega_p^3$, which then satisfies 
\begin{equation}
j^*  = \overline{\overline{v^*}} \> \left(V^*,T^*\right),
\label{sim}
\end{equation}
where $\overline {\overline{v^*}}$ is the reduced velocity averaged over {\it all} the particles.
The reduced average velocity, in turn, must be a function of the two previously introduced control 
parameters: the reduced voltage and temperature. 
Eq.(\ref{sim}) is in fact a similarity transformation relating systems with different charge, length, temperature, and potential difference. 
In  Fig. \ref{fig2.eps} we plot $j^*$ vs. $V^*$ for various $T^*$. The phase diagram provides all the information about the
current-voltage characteristics for all possible planar diodes.  
The first feature to note is that all the different curves emanate from 
the limiting current backbone, which traces a temperature dependent path in the $j^* \times V^*$ plane. 
To the left of the limiting current border, indicated by the solid line in Fig. \ref{fig2.eps}, 
the distribution function can no longer be described by a unidirectional Maxwellian, such as the
expression (\ref{equa3}).  The transition resembles  Bose-Einstein
condensation (BEC).   In the case of BEC --- below the critical temperature --- a macroscopically populated ground 
state appears, and only a  
fraction of  particles remains in the excited states.  Similarly, in the case of our
diode, to the left of the limiting curve,
part of the charge must be expelled from the system before a stationary state can be achieved. As $T_c^* \rightarrow \infty$, 
the voltage effects become negligible compared to the thermal ones, 
and the beam density becomes uniform across the gap. In this limit, 
it is possible to show that the dimensionless backbone curve 
asymptotes to  a vertical line, $V^* = 0.5$, Fig. \ref{fig2.eps}.

%


To conclude, we have studied the dynamics of collisionless driven systems.  
Unlike the stochastic lattice gasses which are significantly abstracted from reality,
the models studied in this paper are very similar to real electronic devises,
such as the thermionic valves, diodes, and magnetrons. Furthermore, differently from the lattice gases
whose dynamics is diffusive, the distribution function of collisionless systems satisfies the Vlasov equation.  For the
class of driven systems introduced in this letter, the  stationary state Vlasov equation can be solved 
exactly.  The theory developed in this paper should, therefore, be
relevant to the design and operation of real electronic devises.  

It is important to stress that  
in the absence of collisions, a charged beam does not relax to an equilibrium with a known 
equation of state.  In fact, the thermodynamic temperature is defined only in the vicinity of the hot emitting cathode.
Away from the cathode,
dynamics is controlled by the collisionless Vlasov equation, which has to be solved as a boundary value problem. 
Once the  solution is obtained, all the
macroscopic quantities can be determined via appropriate averages. The kinetic 
temperature  is found to vary strongly across the device gap, 
precluding the use of conventional isothermal or adiabatic assumptions and of the hydrodynamic formalisms. 

\acknowledgments
This work is supported by CNPq, FAPERGS, INCT-FCx of Brazil, and by the Air Force Office of Scientific Research 
(AFOSR), USA, under the grant FA9550-09-1-0283.

\end{document}